\documentstyle[psfig,aps,prb,multicol]{revtex}

\newcommand{\na}{$\rm NaV_2O_5$ }  
\newcommand{\nae}{$\rm NaV_2O_5$}

\newcommand{\be}{\begin{equation}}
\newcommand{\ee}{\end{equation}}

\newcommand{\no}{\noindent}
\newcommand{\ra}{\rangle}
\newcommand{\et}{{\it et al. }}

\begin{document}

\draft

\title{Orbital-resolved Soft X-Ray Spectroscopy in NaV$_{\bf
2}$O$_{\bf 5}$ }

\author{G. P. Zhang} \address{Department of Physics and Astronomy, The
University of Tennessee, Knoxville, TN 37996-1200 \\ Department of
Physics, State University of New York, College at Buffalo, Buffalo, NY
14222$^*$}

\author{G. T. Woods$^{1,2}$ and Eric L. Shirley$^{2}$} 

\address{$^1$Department of Physics and Astronomy, The University of
Tennessee, Knoxville, TN 37996-1200 \\$^2$ Optical Technology Division,
National Institute of Standards and Technology, \\  Gaithersburg, MD
20899-8441$^\#$ }

\author{T. A. Callcott, L. Lin and G. S. Chang}

\address{Department of Physics and Astronomy, The University of
Tennessee, Knoxville, TN 37996-1200}

\author{B. C.  Sales$^1$, D. Mandrus$^{1,2}$ and J. He$^{1,2}$}

\address{$^1$Solid State Division, Oak Ridge National Laboratory, TN
37831, \\ $^2$Department of Physics and Astronomy, The University of
Tennessee, Knoxville, TN 37996-1200}

\date{\today}
\maketitle

\begin{abstract}
We demonstrate that angle-resolved soft x-ray spectroscopy can resolve
absorption by inequivalent oxygen sites and by different orbitals
belonging to the same site in \nae. By rotating the polarization
direction, we see a dramatic change in the absorption spectra at the
oxygen $K$ edge. Our theory identifies the detailed composition of the
spectra and predicts a correct energy-ordering of the orbitals of
three inequivalent oxygen atoms. Because different orbitals dominate
absorption spectra at different energies and angles, one can excite at
a specific site and ``orbital''.  In contrast, absorption at the
vanadium L edge does not show large changes when varying the
polarization direction.  The reason for this is that different
excitation channels (involving different initial states for the
excited electron) overlap in energy and vary in compensating ways,
obscuring each channel's sensitive polarization dependence.
\end{abstract}




\section{Introduction}

Orbital degrees of freedom play a crucial role in many technologically
interesting materials, such as superconductors, magnetic materials, and
other strongly correlated compounds. For instance, the spin-orbit
coupling effects determine whether the magnetic crystalline anisotropy
in magnetic thin films favors magnetic moments normal or parallel to the
surface of the sample.  Such important effects motivate many
investigations that probe orbitals in real materials. Soft x-ray
spectroscopy (SXS) \cite{tom1,tom2,tom3,tom4,kotani} has a unique
capability to reveal spin and orbital information specific to one
element.  By employing x-ray magnetic circular dichroism, one can
measure spin and orbital moments;\cite{carra} using x-ray magnetic
linear dichroism, one can image antiferromagnetic domains at
sub-micrometer scales.\cite{stohr} Despite these tremendous
achievements, imaging atomic orbitals has not often been done. The first
attempt at the C $K$ edge of graphite was done by Rosenberg {\it et
al.}\cite{rose} A more elaborate study was performed a decade later with
an intense synchrotron source.\cite{sky} On surfaces, the first
investigation was made by Nilsson {\it et al.}\cite{nilsson} They used
$s$-polarized soft x-rays to selectively excite $\pi$ orbitals of CO
adsorbed on Ni and used $p$-polarized light at off-normal angles to
excite both $\sigma$ and $\pi$ orbitals, from which the orbital
information could be extracted. However, obtaining such valuable orbital
information in {\it bulk materials} is still a challenge for SXS.

In this work, we consider \nae, a fascinating transition-metal compound
that features three inequivalent oxygen sites: O$_1$, O$_2$ and O$_3$
[cf Fig. 1(a)].  At 34 K, this material undergoes a spin-Peierls-like
phase transition, which has attracted great
interest.\cite{smo,japan,lee,valenti} \na is orthorhombic with symmetry
of $ P_{mmn}$ and lattice constants $a$=11.316 $\rm \AA$, $b$=3.611 $\rm
\AA$ and $c$=4.797 $\rm \AA$.  It is a ladder structure compound along
the $b$ axis with V-O$_1$-V bonds as its rungs extending along the $a$
axis. The neighboring ladders along the $b$ axis are connected by
shifting a half-unit cell along the $b$ axis. These ladders form
quasi-two dimensional layers between which the sodium atoms are
intercalated.  One O$_1$, three oxygen O$_2$ and one apical O$_3$ atoms
form a pyramid inside which one vanadium atom is situated.  These three
different oxygen atoms present a challenge to site-specific SXS, because
one desires to obtain information, not only about different orbitals of
atoms of different elements, but also about different orbitals on the
same site.

Specifically, we report orbital-resolved x-ray absorption measurements,
using the total electron yield method (TEY),
on \na at the O $K$ (1s) and V $L_3$ (2p$_{3/2}$) edges.  By changing
the angle $\theta$ between the electric field-polarization direction
and certain crystal axes of \nae, we can determine characteristic
orbital energies, not only for orbitals on inequivalent O sites, but
also for different orbitals on the same O site.  With the aid of
theoretical calculations, we are able to pinpoint the origin of the
angular dependence of the O $K$ edge absorption spectrum, distinguish
different contributions thereto, and accurately determine the crucial
energy ordering of the different orbitals from three inequivalent
oxygen atoms. Combining different scans along different axes, we are
able to project out spatial orientations of oxygen orbitals in real
space.  Thus, our study highlights an important capability of SXS that
has not been explored in great detail. \cite{polar}

Our experiment reveals a surprising difference between the vanadium and
oxygen edges. At the V $L_{3}$ edge, the spectra show a very small
dependence on $\theta$, but at the O $K$ edge there is a large
dependence.  Our calculations are scalar-relativistic, so that we model
the V $L_{2,3}$ manifold without spin-orbit effects (including
$p_{1/2}-p_{3/2}$ splitting) taken into account.  However, this
description should still be able to address many aspects of a
polarization dependence, which depends on states near the Fermi level.
The theory shows that, subject to the dipole selection rule $\Delta
l=\pm 1$ and $\Delta m=0,\pm 1$, transitions from $s$ orbitals only have
one initial state, i.e., $m=0$, while transitions from $p$ orbitals have
three initial states, i.e., $m=0, \pm 1 $.  Polarization dependences for
absorption by electrons in each of the initial states compensate each
other and ultimately lead to a very small polarization dependence. This
explains the observed trend for the V $L_3$ (2p$_{3/2}$) edge.

The remainder of this article is arranged as follows.  A brief review of
our experimental measurements is described in Section II.  In Section
III, we present our theoretical scheme. The results and discussions are
shown in Sec. IV. Finally, we conclude our paper in Section V.  In a
companion paper,\cite{gw} we analyze emission spectra and
experimentally determine the occupied $p$ density of states from each
oxygen site. This paper is devoted entirely to the analysis of
absorption spectra.

\section{Experimental measurements}

Single crystals of \na were prepared at the Oak Ridge National
Laboratory.\cite{gw} The crystals grow as rectangular platelets with the
c axis normal to the plate surface, the a axis along the short side of
the rectangle and b along the long side, as shown in Fig. 1(b).  The
samples were cleaved along the c axis. The experimental measurements
were done at Beamline 8.0 of the Advanced Light Source at the Ernest
Orlando Lawrence Berkeley National Laboratory.\cite{gw} The geometry of
the measurements is indicated in Fig. 1(b). Horizontally polarized
x-rays from an undulator and monochromator were incident on the sample.
In the TEY detection mode, the x-ray absorption spectrum is measured by
collecting the electrons that escape from the sample.  Two samples were
used in the experiment, and mounted at right angles to each
other. Rotation of the sample holders scanned the polarization vector of
the incident x-rays in the a-c plane for one sample and in b-c plane for
the second sample.

\na is an ideal system, in which to reveal a polarization dependence,
because its structure is highly anisotropic.  In Fig. 2, we show the
normalized absorption spectra of a polarization scan from along the b
axis to along the c axis, specifically from $\theta=7.5^{\circ}$ to
$\theta=75^{\circ}$ with a step of 7.5$^{\circ}$.  For each angle, the
normalization is done with respect to the total area between 514 eV and
536 eV because this energy region covers both the V-L and O-K edges.

We note that the TEY method has a finite sampling depth of about
100~$\rm \AA$.  This suggests that our electron yield arises from a
local spot of the sample. To illuminate the sample in the same place
during measurement and to ensure data consistency, we adjusted the
sample position after each change of its orientation angle.  We found
that data obtained using this technique are reproducible within 5~\%.

We first consider the vanadium $L_{3}$ edge around 518 eV
[Fig. 2(a)]. When we scan the polarization from along the b axis to
along the c axis, the spectra change very little.  We observed a similar
behavior when we scanned from along the a axis to along the c axis (not
shown).  However, for the oxygen $K$ edge, we see a totally different
picture [Fig. 2 (b)].  Increasing the angle $\theta$ from 7.5$^{\circ}$
to 75$^{\circ}$ strongly changes the spectra. Three main peaks A, B and
C, located at 530.6 eV, 532.1 eV and 532.8 eV, respectively, can be
clearly identified.  Peak A's intensity decreases sharply and eventually
disappears as the polarization moves to along the c axis, which is
normal to the sample surface.  In the meantime, intensities of peaks B
and C increase strongly and become the two main peaks on the spectra. To
have a clear view of the intensity change, in Fig. 3 we plot the
intensities versus angle $\theta$ for those three peaks.  Their
variations can be nicely fitted to a $\cos 2\theta$ dependence, which is
the simplest possible variation (see the solid lines in Fig. 3).

\newcommand{\vrr}{ {\bf r}}

\section{Theoretical framework}

Our theoretical calculation to simulate the experiment is based on
pseudopotential,\cite{Pickett} local-density-approximation (LDA)
calculations.\cite{KohnSham} Electron states are found by solving the
self-consistent Kohn-Sham (KS) equations, \be \hat{H}\Psi_{n{\bf
k}}({\bf r})=\{ -\frac{{\hbar}^2}{2m}{\nabla}^2+ \hat{V}_{\rm eff}\}
\Psi_{n{\bf k}}({\bf r})=E_{n{\bf k}}\Psi_{n{\bf k}}({\bf r}) \nonumber
\ee
\noindent
Here $\Psi_{n{\bf k}}$ is the Bloch state for an electron in band $n$
at crystal wave-vector ${\bf k}$.
The effective potential is given by
\be
\hat{V}_{\rm eff}={V}_{\rm ion}+
V_{\rm H}(\vrr)+V_{\rm xc}(\vrr).
\ee
\noindent
The three terms account for the pseudopotential interaction of electrons
with ion cores $V_{\rm ion}$, the Hartree potential $V_{\rm H}({\bf r})$
that is obtained from the electron charge density, given by \be
n(\vrr)=\sum_{n{\bf k}}^{\rm occ'd.} |\Psi_{n{\bf k}}(\vrr)|^2 \ee and
exchange and correlation effects $V_{\rm xc}$.  We use the
Ceperley-Alder form\cite{capz} for the exchange term $V_{\rm xc}={\delta
E_{\rm xc}[n]}/{\delta n({\vrr})}$. The ionic potential $V_{\rm ion}$ is
a norm-conserving pseudopotential computed within the Hartree-Fock
approximation, and with core-valence correlation effects included using
the core-polarization-potential method.\cite{eric1} As basis sets, we
use plane waves or optimized basis functions.\cite{eric2}

To compute an x-ray absorption spectrum, we use the
expression,\cite{tom1,kotani} \be S(\omega)\propto 2\pi \sum_f|\langle
f|{\bf p}\cdot {\bf A}|i\rangle|^2 ~\delta(\omega+E_i-E_f) \ee \no where
${\bf p}\cdot {\bf A}$ is the electron-photon interaction approximated
by a dipole operator; $|i\ra$ is the initial core state with energy
$E_i$, $|f\ra$ is the final state with energy $E_f$, and $\omega$ is the
incident photon energy.  In our present analysis, which mainly deals
with determining the symmetries of orbitals, we neglect electron-core
hole interactions in the final state.

\section{Results and discussions}

\subsection{Oxygen $K$ (1s) edge absorption spectrum}

In the experiment, we always rotated our polarization direction from
along the b axis to along the c axis in the way already described.
Theoretically, we did calculations in the same way.  In Fig. 4(a), we
show theoretical x-ray absorption spectra at the O $K$ edge, where three
peaks, A, B and C, are also clearly observed with different intensities
as a function of polarization. The spectra were calculated with
$12\times 12 \times 12$ k-point meshes. The results are found
well-converged.  The angular dependence of these peaks is in good
agreement with the experimental observations: Peak A decreases as the
polarization moves away from along the b axis, whereas both peaks B and
C increase. The change also follows a cosine function of $\theta$,
consistent with the experimental observation.  Note that the
experimental measurements cannot directly determine which oxygen
orbitals contribute to these main peaks, whereas calculations should
provide important insight into this.

Figures 4(b-d) show the respective contributions from three
inequivalent oxygen sites.  To simplify our notation, we let
O$_m$(P$_n$) denote the $n$th orbital of oxygen O$_m$ that
participates the absorption process.  That is, the local, partial
density of states for such an orbital can play a significant role in
absorption as a function of excitation energy, and this role can be
strongest at some ``characteristic'' energy.  We first consider peaks
B and C.  By comparing Fig. 4(a) with Figs. 4(b), 4(c) and 4(d), we
infer that peaks B and C in Fig. 4(a) and Fig. 2(b) originate from
O$_1$(P$_3$), O$_2$(P$_4$), O$_2$(P$_5$), and O$_3$(P$_5$) orbitals,
whose characteristic orbital energies (expressed relative to the O 1s
level) are 533.0~eV, 533.1~eV, 533.7~eV and 533.1~eV, respectively.

When considering peak A in Figs. 4(a) and 2(b), we notice that all
three inequivalent atoms contribute but are favored in different
ranges of energy.  Figure 4(b) illustrates the main contributions from
O$_1$. It is important to note that only O$_1$(P$_1$) appears at 530.2
eV, which helps us to resolve different orbital contributions (see
next paragraph).  The peak intensity decreases as the polarization
direction deviates from the b axis, in a similar way as peak A in
Fig. 2(b) or Fig. 4(a). O$_1$(P$_1$) is immediately followed by
O$_3$(P$_1$) at 530.6 eV with an energy difference between them of
about 0.4~eV, [cf. Fig. 4 (d)]. The O$_3$(P$_1$) intensity changes in
a similar manner as does that for O$_1$(P$_1$).  After the
O$_3$(P$_1$) orbital, the O$_2$(P$_3$) orbital gradually appears
around 531~eV to 532~eV [cf. Fig. 4(c)].  Because of their unique
positions, O$_2$ orbitals are distorted, and not along one particular
axis.  Thus, the angular dependence of their contribution to
absorption is weaker than for orbitals on O$_1$ and O$_3$.

These results show that the measured peak A has contributions from the
O$_1$(P$_1$), O$_3$(P$_1$), and part of O$_2$(P$_3$) orbitals.  This
could prove crucial for our tentative explanation of emission spectra
that is given subsequently in Ref. 16.  In particular, because our
theory shows that these three orbitals have different energies, it is
possible to excite different orbitals at different energies and to {\it
resolve} them experimentally. This is exactly what we did.  Because the
energy increases in the order of O$_1$, O$_3$ and O$_2$, we can
selectively excite these atoms.  When we excite at the O$_1$(P$_1$) {\it
orbital} edge, we only see one peak in the emission spectrum, because
O$_2$ and O$_3$ atoms are not excited, though in the normal fluorescence
spectrum, we expect to see two main peaks.  Therefore, with theoretical
guidance, we demonstrate that we are able to {\it resolve} these {\it
different} orbitals associated not only with different sites such as
O$_1$(P$_1$), O$_3$(P$_1$), O$_2$(P$_3$) but also with the {\it same}
site such as O$_1$(P$_1$) and O$_1$(P$_3$).

By scanning polarization from along the a axis to along the b axis 
and from along the a axis to along the c axis, we can project
out orbital orientations in real space.  Putting together all the
results, we can show some of the orbitals in the real space in Fig. 5.
One sees that there are two kinds of O$_3$ orbitals: one is along the
a axis, the other is along the c axis, and one O$_1$ orbital along the
b axis, two O$_2$ orbitals along the b and a axes.  Note that,
unlike charge-density plots for orbitals,
the orbitals presented here are experimentally accessible.  Thus, we
demonstrate that the soft x-ray spectroscopy has the capability of
resolving orbitals from the same site.

\subsection {Vanadium $L_3$ (2p$_{3/2}$) edge}

The x-ray absorption spectrum at the V $L_3$ edge is rather unique and
shows a very weak angular dependence. At room temperature, there is only
one inequivalent vanadium atom.  In our scalar-relativistic
calculations, excitation of V 2p electrons can involve three electron
initial states: $m=0, \pm 1$.  Our absorption spectrum intrinsically
involves a sum over excitations with the indicated initial states, and
the experimentally observed spectrum is an analogous sum at the V $L_3$
edge.  We first show the results of the first channel in Fig. 6(a). We
note that with the incident polarization turning away from the b axis,
the whole spectrum clearly changes. If this were the dominant excitation
channel, then one would see a strong angular dependence in
experiment. However, our experiment showed otherwise, which implies that
we should take into account the other channels. In Fig. 6(b), we plot
the change because of the $l=1$, $m=\pm1$ channel.  Without spin-orbit
coupling, the spectra for both $l=1,m=+1$ and $l=1,m=-1$ channels are
same.  Two prominent features are: (1) the mean peak strongly overlaps
the spectra of the $l=1$, $m=0$ channel; (2) the portion showing strong
angular dependence is on the right-hand side, which is opposite to that
in Fig. 6(a). Because of this compensating effect, the total absorption
spectrum does not show a strong angular dependence as one can see in
Fig. 6(c). This explains our experimental observations.

Now comparing this with the oxygen $K$-edge spectra where only one
excitation channel, $l=0$ and $m=0$, participates in the process, we
clearly see that appropriate summation over the multiple excitation
channels (i.e., corresponding to different electron initial states)
obscures the polarization dependence, which ultimately leads to the
big difference between the role of polarization at the V $L_3$ edge vs
the O $K$ edge. This finding demonstrates the importance of
understanding selection rules in x-ray spectroscopy, and it might help
one optimize the polarization direction to reveal the maximal
information from x-ray spectroscopy measurements in the future. In
particular, it would be interesting to ask whether the present
polarization effects for the various components of the V $L_3$
spectrum could be resolved by circular dichroism. However, the
difficulty is that even with the 2p core spin-orbit coupling, in the
absence of spin-polarization of the V-d bands, $\Delta m=\pm 1$
transitions should also be the same.

\section {Conclusions}

In conclusion, we measured angular dependence of x-ray absorption
spectra at the vanadium $L_3$ edge and oxygen $K$ edge. Rotating the
polarization from along the b axis to along the c axis, the spectra at
these two edges behave quite differently. There is very little change at
the V $L_3$ edge, while a dramatic change is observed at the O $K$
edge. Such a large change enables us to selectively excite different
orbitals belonging to the same site. Thus, we are able to {\it resolve
orbitals} for the {\it same} element using soft x-ray spectroscopy. Our
theoretical calculation provides the detailed information to understand
the experimental spectra.  It clarifies how the predominant first peak
at the oxygen $K$ edge is from three different orbitals of three
different oxygen sites (O$_1$, O$_2$ and O$_3$) and determines the
crucial energy ordering of the orbitals on these sites.  This also helps
explain the emission spectra.  Our calculations also show how a
compensating effect can arise in the polarization dependence of various
excitation channels for the V $L_3$ edge, so that the total absorption
spectrum at the edges exhibits a small polarization dependence.  After
we submitted this paper, we noticed three recent papers \cite{three}
that intensively debated whether the anomalous scattering factors in the
x-ray diffraction technique could be used to probe the charge ordering
in \nae. Our study shows that if excited at the V $L_3$ edge, the
orbital or structural effects are small. Thus, one might explain the
anomalous scattering as resulting from the charge ordering, provided
that the scattering intensity of the charge ordering is larger than
that of the orbital ordering.

\noindent {\bf Acknowledgment}

This research is supported by NSF DMR-9801804 and NSF DMR-0072998. Oak
Ridge National Laboratory is managed by UT-Battelle, LLC, for the
U.S. Department of Energy under Contract No. DE-AC05-00OR22725.
Measurements were carried out at the Advanced Light Source at the Ernest
Orlando Lawrence Berkeley National Laboratory, supported U.S. Department
of Energy Contract DE-A003-76SF00098. We would like to acknowledge
Christen Halloy and the staff at the Joint Institute of Computational
Sciences (JICS) at the University of Tennessee at Knoxville, where part
of calculation has been done.

\noindent $^*$Mailing address

\noindent $^\#$Present address

\newpage

\begin{figure}
\caption{ (a) \na structure; (b) Experimental geometry. }
\end{figure}

\begin{figure}
\caption{ Experimental x-ray absorption spectra of \na at (a) the V
$L_3$ edge and (b) the O $K$ edge.  The angle $\theta$ is between the
electric polarization and the b axis. $\theta$ increases from
$7.5^{\circ}$ to $75^{\circ}$ in steps of 7.5$^{\circ}$. The scan is
in the b-c plane. The vertical dashed lines denote the positions of
peaks A, B and C at 530.6, 532.1 and 532.8 eV, respectively. }
\end{figure}

\begin{figure}
\caption{Intensities versus the polarization angle $\theta$ for peaks
A, B and C in Fig. 2(b).}
\end{figure}

\begin{figure}
\caption{ Theoretical absorption spectra of \na at the O $K$ edge.
(a) Total absorption spectra at the O $K$ edge.  Contributions are
from (b) O$_1$, (c) O$_2$ and (d) O$_3$.  The scan condition is same
as in Fig. 2. The arrows denote the change in intensity while varying
polarization from along the b axis to along the c axis.}
\end{figure}

\begin{figure}
\caption{Oxygen orbitals' spatial distributions in \nae.}
\end{figure}

\begin{figure}
\caption{Theoretical V $L_{3}$ edge absorption.  (a) Only the channels
$l=1$ and $m=0$ is calculated; (b) Only the channel $l=1$ and $m=\pm1$
are calculated. (c) Total absorption spectrum at the V $L_{3}$
edge. The scan condition is exactly same as in Fig. 2. The arrows have
the same meaning as in Fig. 4.}
\end{figure}
\newpage

~~~~~~~~~~~~~~
\vspace{1cm}

\hspace{0cm}\psfig{figure=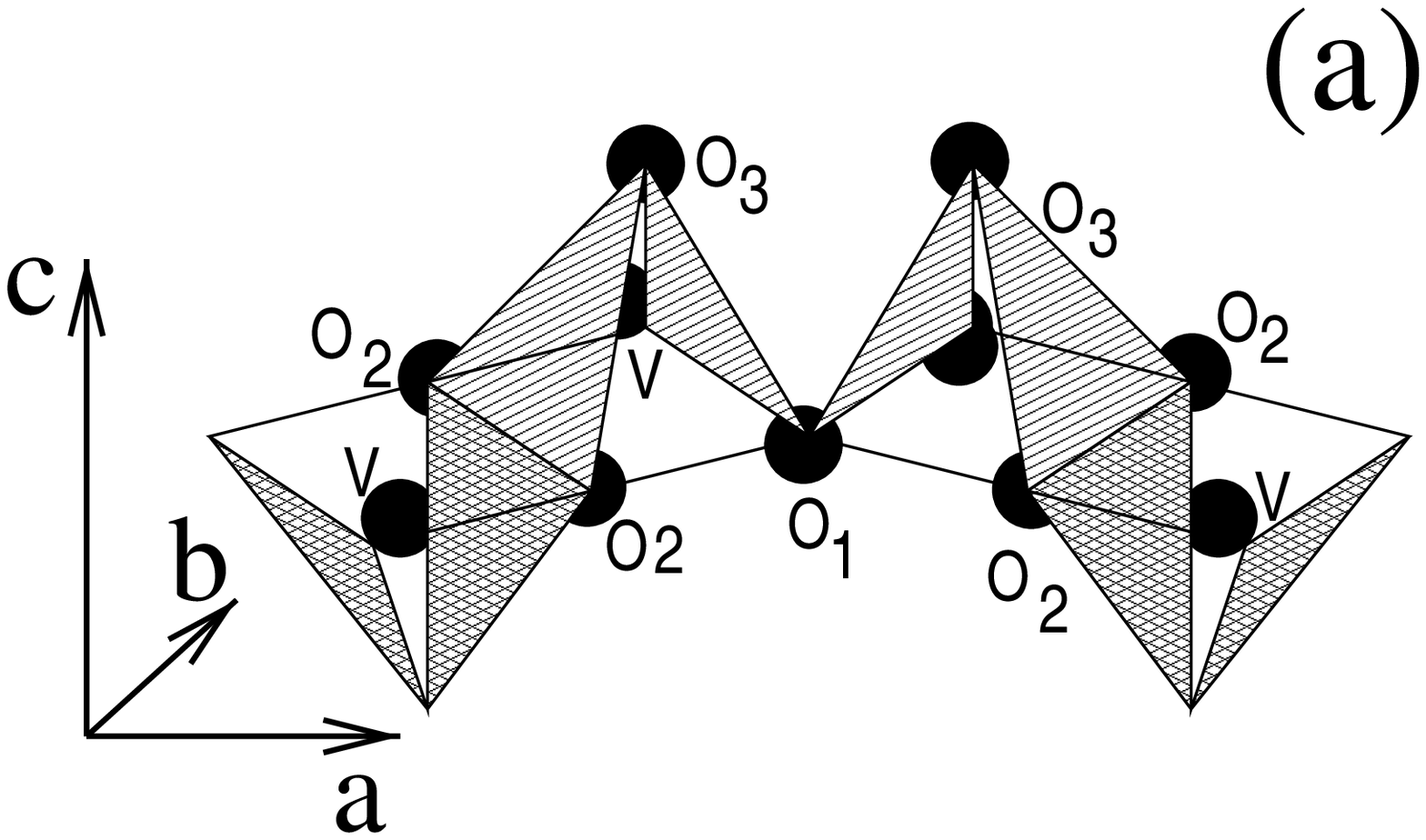,width=13cm,angle=0}

\vspace{1cm}

\hspace{2cm}\psfig{figure=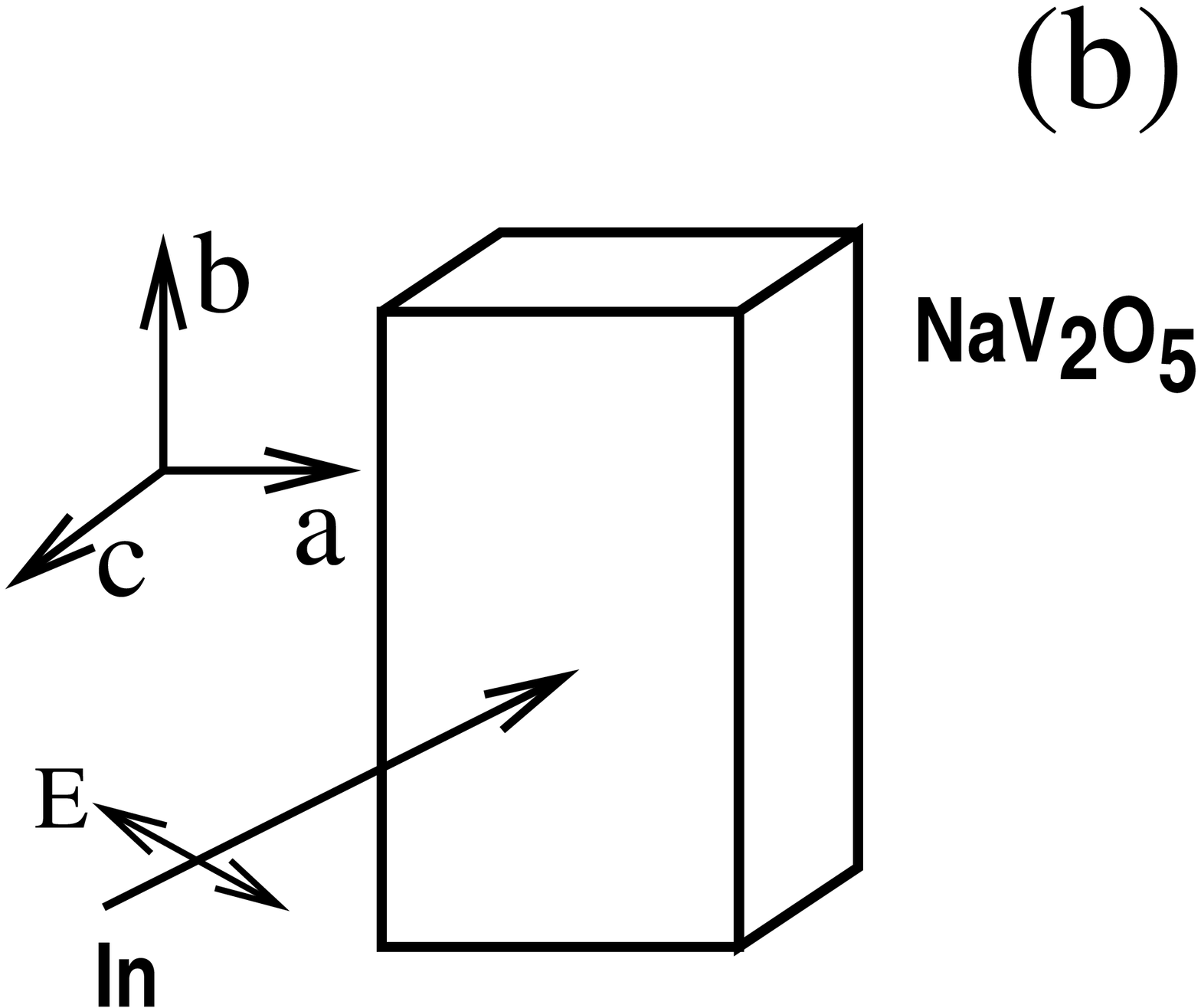,width=11cm,angle=0}

\vspace{1cm}
\centerline{Figure 1}

\newpage

~~~~~~~~~~~~~~
\vspace{2cm}

\hspace{2cm}\psfig{figure=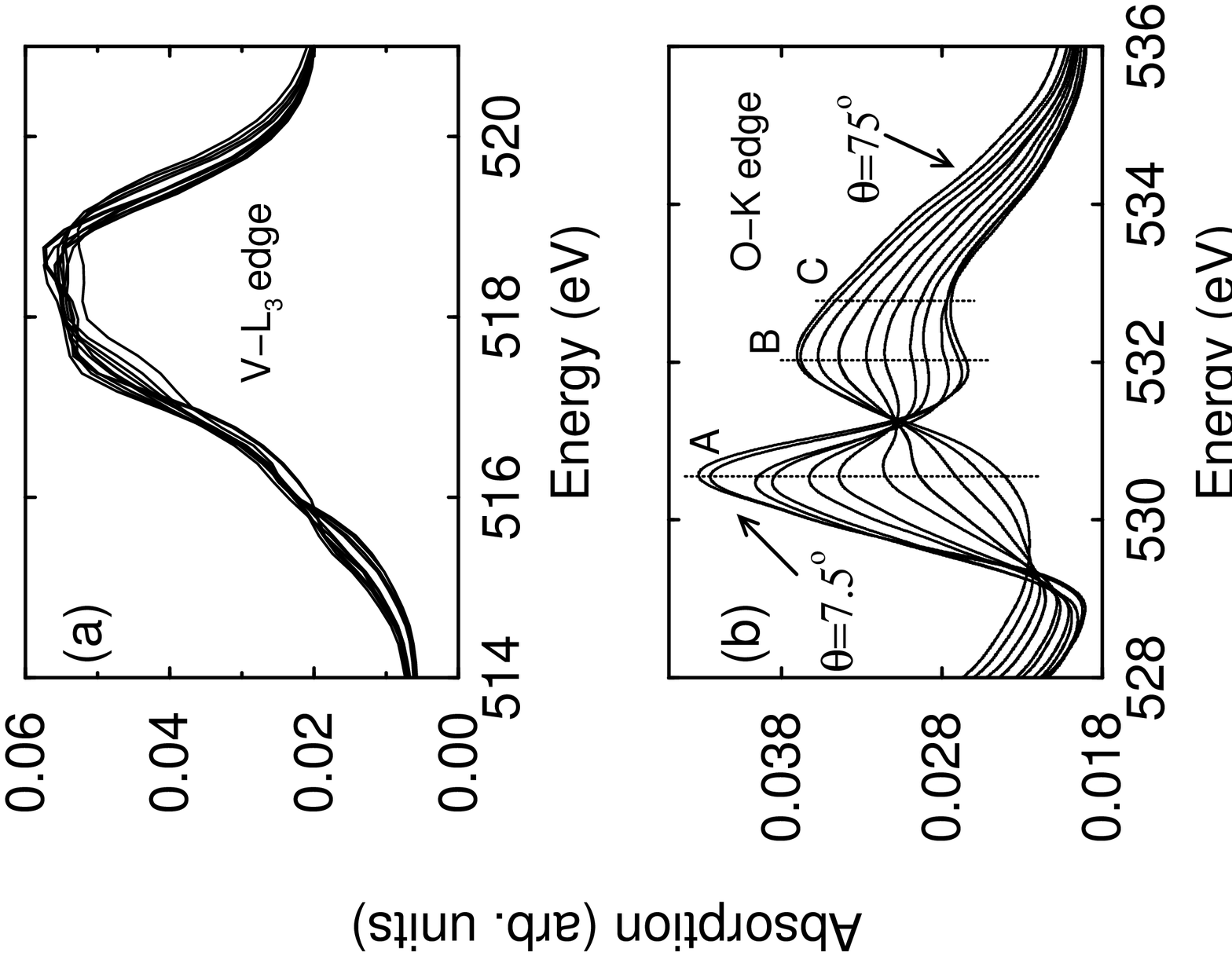,width=10cm,angle=270}

\vspace{5cm}
\centerline{Figure 2}

\newpage
~~~~~~~~~~~~~~
\vspace{1cm}

\hspace{2cm}\psfig{figure=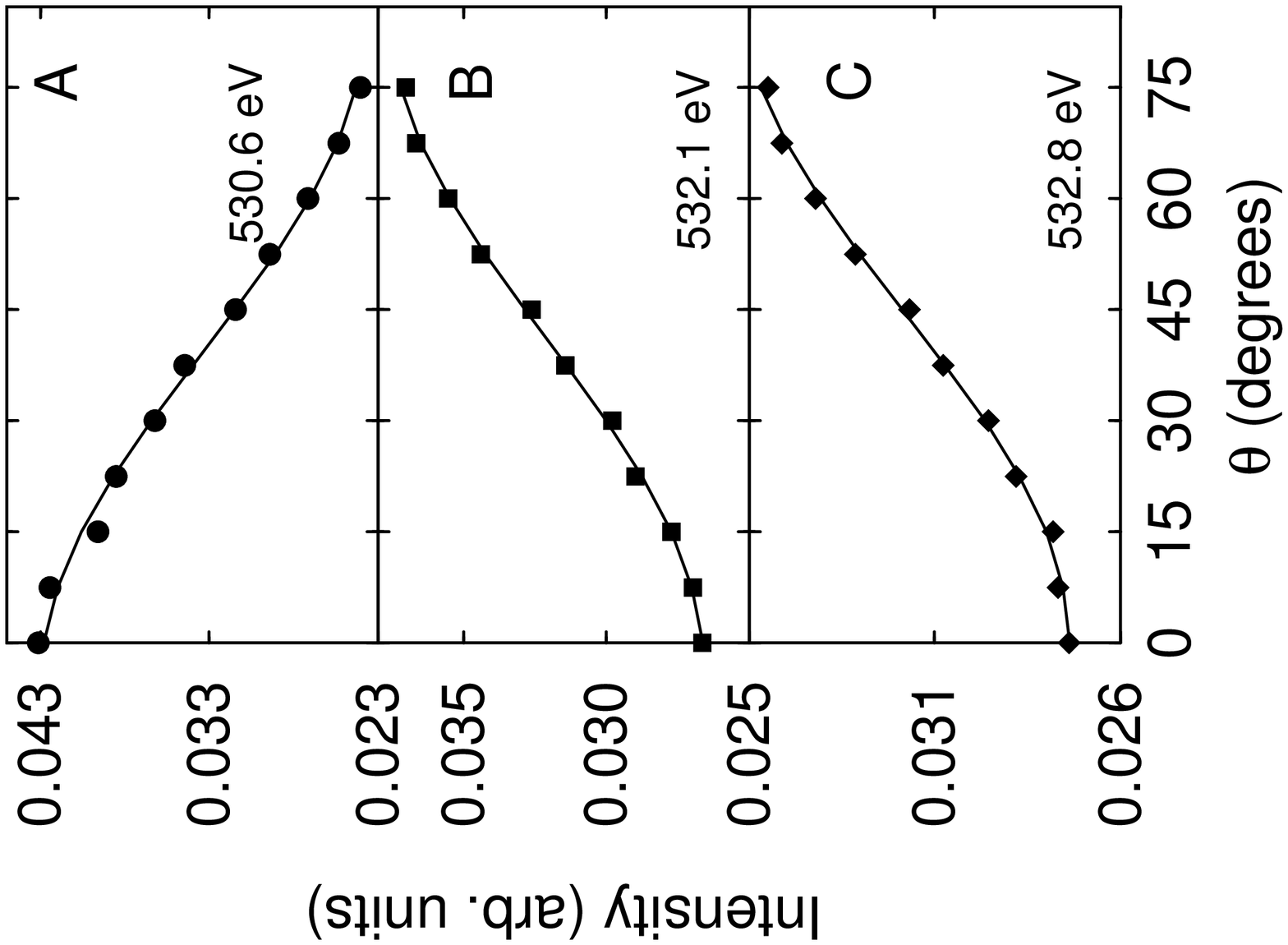,width=9cm,angle=270}

\vspace{5cm}
\centerline{Figure 3}

\newpage
~~~~~~~~~~~~~~
\vspace{2cm}

\hspace{2cm}\psfig{figure=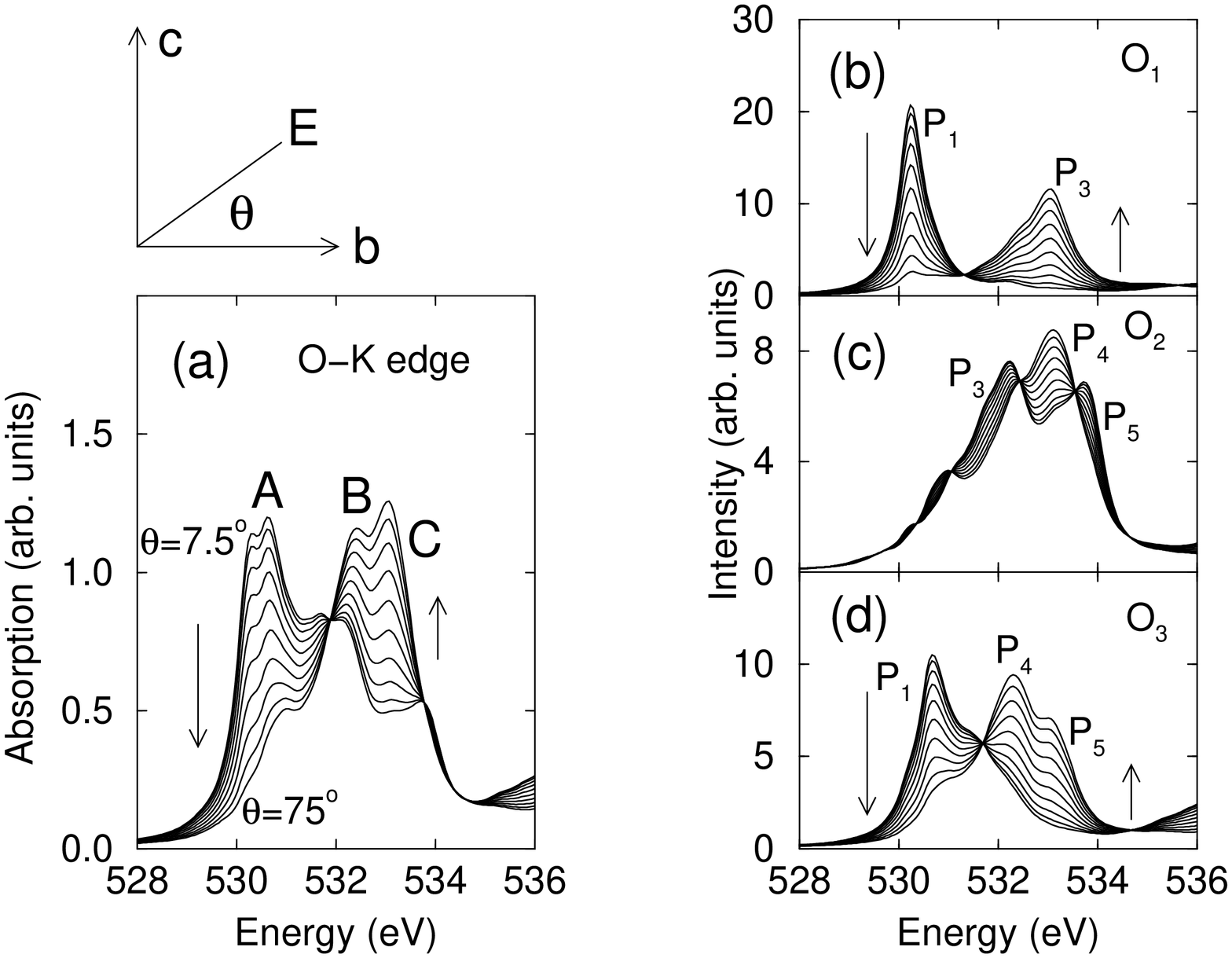,width=16cm,angle=270}

\vspace{5cm}
\centerline{Figure 4}

\newpage
~~~~~~~~~~~~~~
\vspace{2cm}

\hspace{2cm}\psfig{figure=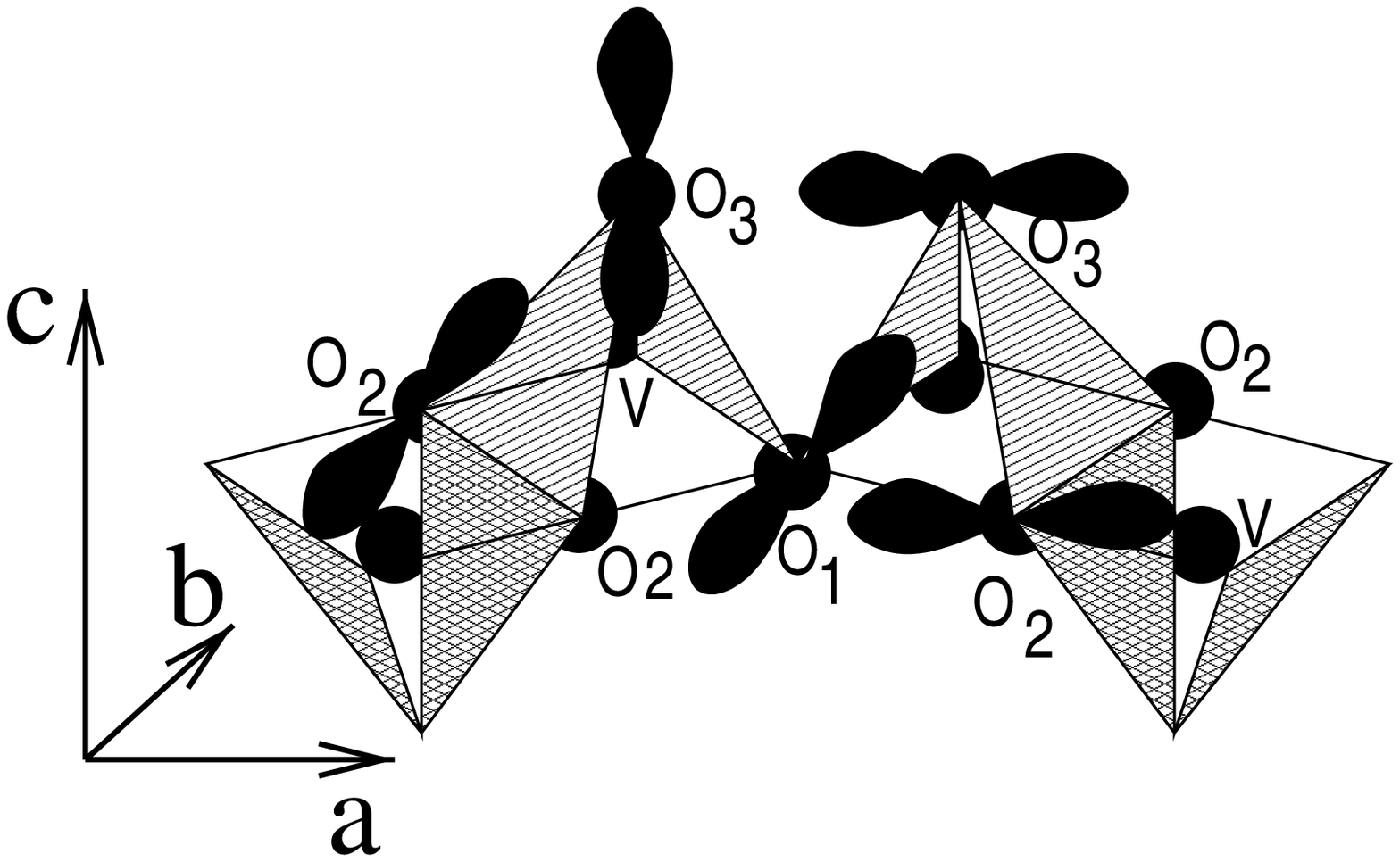,width=14cm,angle=0}

\vspace{8cm}
\centerline{Figure 5}

\newpage
~~~~~~~~~~~~~~
\vspace{2cm}

\hspace{4.5cm}\psfig{figure=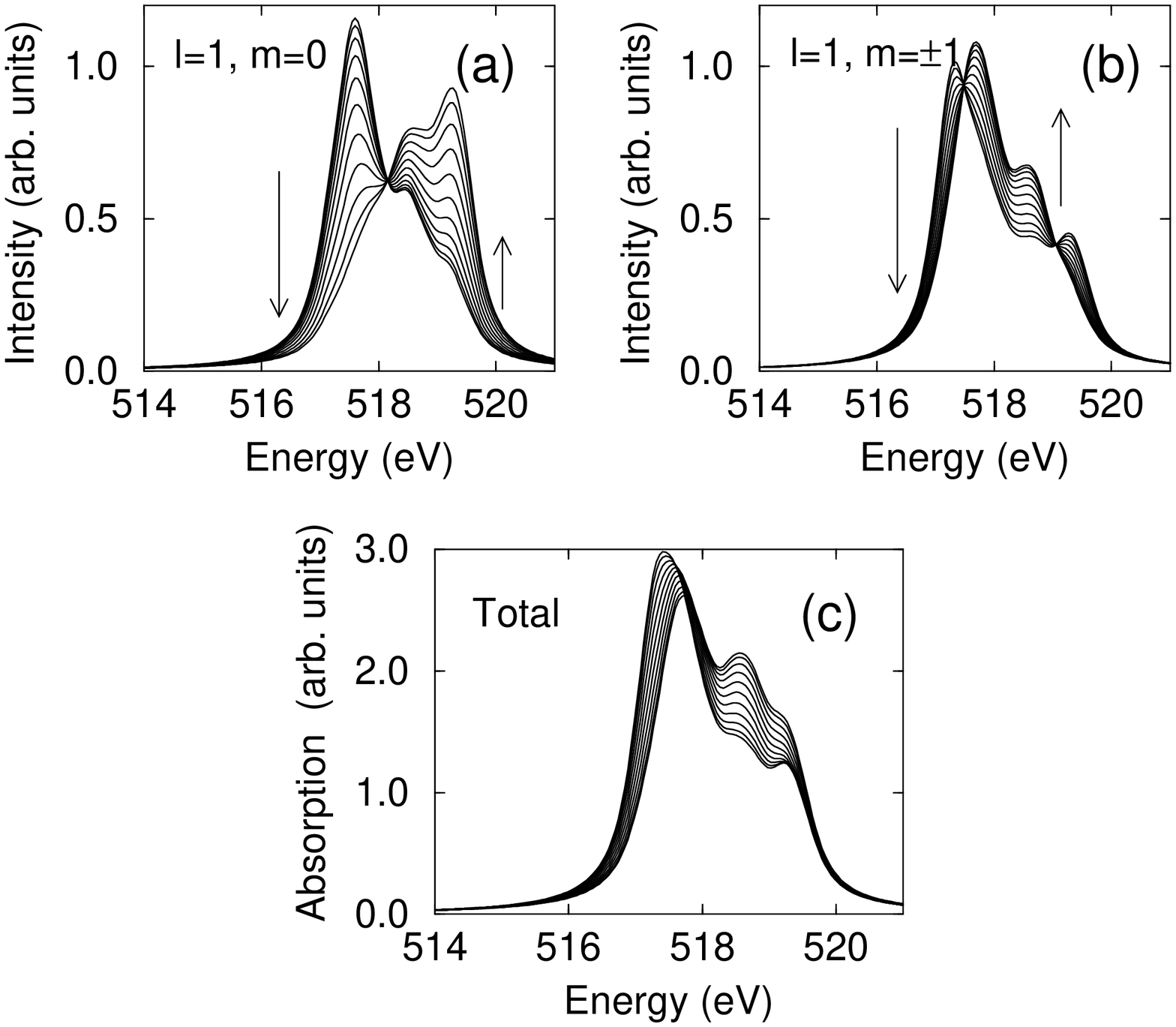,width=14cm,angle=270}

\vspace{4cm}
\centerline{Figure 6}

\end{document}